# New thickness and shape parameters for describing the thermal boundary layer

David Weyburne
Air Force Research Laboratory, 2241 Avionics Circle, WPAFB, OH 45433, USA
E-mail: David.Weyburne@gmail.com

**Abstract**
New thickness and shape parameters for describing the thermal boundary layer formed by 2-D fluid flow along a heated (cooled) wall are presented. The parameters are based on probability density function methodology in which the thermal profile thickness and shape are described by central moments of the thermal profile. The moment-based parameters are simple integrals of the temperature profile. The usefulness of the new method is demonstrated by applying the description to laminar and turbulent boundary layer datasets.

## 1. Introduction

Prandtl[1] introduced the concept a boundary layer for fluid flow past a solid over a hundred years ago. The boundary layer concept for flow over a flat plate is depicted in Fig. 1. As the flowing fluid impinges on the flat plate a boundary layer is formed along the plate such that the velocity and temperature at the wall gradually transition to the bulk values. In spite of the vast amount of development in the field, the mathematical tools for describing the temperature profiles that form along the plate are still very limited. Recently Weyburne[2,3] introduced the moment method borrowed from probability theory to describe the velocity profile formed by 2-D fluid flow along a wall. The method was developed from the observation that for laminar flow over a flat plate, the second derivative of the velocity and the temperature in the direction normal to the plate has a Gaussian-like profile.[2] In the more recent work,[3] the moment method for the velocity profile was expanded by adding descriptive parameters based on moments of the first derivative and the velocity profile itself. Borrowing from probability density function methodology, the boundary layer is described in terms of the central moments of these Gaussian-like kernels. The most important result of these efforts is a physically meaningful set of parameters that describe the boundary layer thickness and shape.

In the earlier paper,[2] the temperature profile formed by the fluid flowing over a heated wall was also considered using the same moment-based method. However, only the second derivative descriptive moments were considered. As we demonstrate below, the second derivative moments basically tell us the thickness and shape of the near wall region of the boundary layer where thermal diffusivity-based heat transfer is important. However, for the case of the turbulent boundary layer, this does not fully define the extent of the thermal boundary layer. In the work described herein, we explore other integral kernels and find the kernel based on the temperature profile and a second one based on its first derivative works well for this purpose. To demonstrate the utility of the new descriptive tools, experimental data for laminar and turbulent boundary layers are examined in the Sections below.

In what follows, we begin by reviewing the original thermal boundary layer description that was based on the second derivative of the temperature profile. The two new temperature profile-based moment descriptions are then introduced. Experimental datasets are then used to demonstrate the utility of the new descriptive approach.

## 2. Thermal Boundary Layer Second Derivative Moments

The intention herein is to develop mathematical tools to describe the thickness and shape of the temperature profile formed by fluid flow along a heated/cooled plate. The original thermal boundary layer description[2] was developed from the observation that the second derivative of the temperature profile for laminar flow along a plate had a Gaussian curve appearance.

As confirmation, consider the following. If one starts by assuming that the second derivative of the temperature profile is a Gaussian curve, then one can integrate it twice with the appropriate boundary conditions to find an approximate temperature profile. In this case, one finds that the reduced temperature, $\theta(y)=(T(y)-T_\infty)/(T_w-T_\infty)$, ($T_w$ and $T_\infty$ defined in Fig. 1) at a height $y$ above the wall is approximated very well as

$$\theta(y) \cong 1+\left(\frac{\sigma_d}{z\gamma_1\sqrt{\frac{\pi}{2}}}e^{-\frac{1}{2}\left(\frac{y-\gamma_1}{\sigma_d}\right)^2}-\frac{\sigma_d}{z\gamma_1\sqrt{\frac{\pi}{2}}}e^{-\frac{\gamma_1^2}{2\sigma_d^2}}\right) \quad (1)$$

$$+\frac{y}{z\gamma_1}\left\{ERF\left(\frac{\sqrt{2}}{2}\left(\frac{y-\gamma_1}{\sigma_d}\right)\right)-1\right\}-\frac{1}{z}\left\{ERF\left(\frac{\sqrt{2}}{2}\left(\frac{y-\gamma_1}{\sigma_d}\right)\right)+ERF\left(\frac{\sqrt{2}\gamma_1}{2\sigma_d}\right)\right\}$$

where $z = 1 + ERF\left(\sqrt{2}\gamma_1/(2\sigma_d)\right) + \sqrt{2}\sigma_1 EXP\left(-(\gamma_1/\sigma_d)^2/2\right)/(\gamma_1\sqrt{\pi})$ and where we have introduced two parameters; $\sigma_d$ the thermal boundary layer width, and $\gamma_1$ the mean location of the thermal boundary layer. In Fig. 2, we have plotted the reduced temperature approximation given by Eq. 1 as the dashed line (the values used for $\sigma_d$ and $\gamma_1$ are given below) along with the Pohlhausen[4] similarity solution to the set of coupled governing equations for fluid flow as the solid line. In this figure the reduced temperatures are plotted versus the Blasius[5] reduced height, $\eta$, given by

$$\eta = y\sqrt{\frac{u_\infty}{v_\infty x}}, \quad (2)$$

where $v_\infty$ is the fluids kinematic viscosity. The solid line is generated with a simple FORTRAN program that uses the Blasius[5] solution for the velocity profile coupled with the nondimensionalized thermal differential equation (see Schlichting's[6] Eq. 12.67). The overlapping curves confirm our assertion that the behavior of the second derivative of the reduced temperature for laminar flow is indeed Gaussian-like.

It is evident that if the temperature profile is Gaussian-like, then the mathematical tools used to describe probability density functions can be employed to describe the temperature profiles thickness and shape. This implies that the thermal boundary layer can be described in terms of the second derivative central moments, $\phi_n$, given as

$$\phi_n \equiv \int_0^\infty dy\,(y-\gamma_1)^n\,\frac{d^2\{\gamma_1\theta(y)\}}{dy^2}, \quad (3)$$

where $\gamma_1$ is the normalizing constant and also happens to be the boundary layer mean location defined as

$$\gamma_1 \equiv \int_0^\infty dy\, y\,\frac{d^2\{\gamma_1\theta(y)\}}{dy^2}. \quad (4)$$

The mean location parameter $\gamma_1$ can be found from the requirement that the zeroeth moment must be unity so that

$$\phi_0 = 1 \equiv \int_0^\infty dy\,\frac{d^2\{\gamma_1\theta(y)\}}{dy^2} \Rightarrow \frac{1}{\gamma_1}=-\left.\frac{d\theta(y)}{dy}\right|_{y=0}. \quad (5)$$

For the laminar flow along a flat plate, the mean location can be shown to reduce to $\gamma_1 = -\sqrt{\nu_\infty x/U_\infty}/\theta'|_{\eta=0}$ where the prime denotes differentiation with respect to $\eta$, the Blasius[5] reduced height parameter. Pohlhausen[4] solved the thermal differential equation using the velocity profile from Blasius[5] and showed that to a good approximation $\theta'|_{\eta=0} \cong -0.332\, Pr^{0.343}$ for 0.6<$Pr$<10 where $Pr$ is the Prandtl number. This allows us to estimate the small temperature difference mean location $\gamma_1$ knowing only the Prandtl number.

If we define the thermal displacement thickness as

$$\delta_T^* = \int_0^\infty dy\, \theta(y) \quad , \tag{6}$$

then the second derivative thermal boundary layer width, $\sigma_d = \sqrt{\phi_2}$, can be shown to reduce to $\sigma_d = \sqrt{-\gamma_1^2 + 2\gamma_1\delta_T^*}$ using integration by parts. This equation coupled with the Pohlhausen based approximation for $\gamma_1$ is used to calculate the approximate temperature profile (dashed line) in Figure 2. Note there are no adjustable parameters in Eq. 1 and the only necessary integral value is $\delta_T^*$, the area under the scaled temperature profile (we used the Pohlhausen $\delta_T^*$ value in the Eq. 1 calculations).

In order to appreciate the usefulness of the second derivative moment method, consider the energy conservation equation for turbulent boundary layer flow given by

$$u\frac{\partial \bar{T}}{\partial x} + v\frac{\partial \bar{T}}{\partial y} = \alpha\frac{\partial^2 \bar{T}}{\partial y^2} - \frac{\partial \overline{\hat{v}\hat{T}}}{\partial y} \quad , \tag{7}$$

where $\alpha$ the thermal diffusivity, and where the standard Reynolds decomposition is used to express the temperatures and velocities in terms of the average velocities $u$ and $v$ and average temperature, $\bar{T}$, and the fluctuating components as $\hat{v}$ and $\hat{T}$. The second derivative term basically determines where in the boundary layer that thermal diffusive heat transfer is predominant. Typically, this region is in the near wall region. Thus, the second derivative moment method gives us a way to determine the thickness and shape of the near wall region where thermal diffusive heat transfer is important. For laminar flow, this constitutes the whole boundary layer. On the other hand, for turbulent boundary layers the second derivative moments indicate this region is in the near wall section of the thermal boundary layer as we demonstrate below in Section 4.

**3. Thermal Boundary Layer Temperature Profile Moments**
**3a. Thermal Profile Moments**

For turbulent boundary layers, we searched for alternative kernels that would work for the whole thermal boundary layer and not just the near wall region. One kernel that is found to work well is the kernel based on the thermal profile itself. The proposition then, is that we can describe the thermal boundary layer in terms of the central moments, $\kappa_n$, given by

$$\kappa_n = \int_0^\infty dy\, (y - \zeta_T)^n \, \frac{\theta(y)}{\delta_T^*} \quad , \tag{8}$$

where the thermal displacement thickness $\delta_T^*$ (Eq. 6) is the normalizing constant. The first moment about zero, which we will call the mean location, is defined as

$$\zeta_T = \int_0^\infty dy\, y \, \frac{\theta(y)}{\delta_T^*} \quad . \tag{9}$$

Using these moment equations, it is possible to describe the thermal boundary profile thickness in terms of the mean location $\zeta_T$, and the thermal boundary layer width, $\sigma_T = \sqrt{\kappa_2}$. For the boundary layer thickness $\delta_T$, we find the four-sigma thickness, defined as $\delta_T = \zeta_T + 4\sigma_T$, behaves similar to the 99% thickness for the temperature profile. Note that the mean location $\zeta_T$ and thermal width $\sigma_T$ will in general, take on different numerical values then the second derivative based $\gamma_1$ and $\sigma_d$ values for turbulent boundary layers.

The physical description of the thermal boundary layer using this alternative kernel can be expanded by defining a pair of shape parameters; the thermal boundary layer skewness, $\chi_T \equiv \kappa_3 / \sigma_T^3$ and the excess, $\xi_T \equiv (\kappa_4 / \sigma_T^4) - 3$. For the thermal excess, we follow standard probability density function practice to include the "-3" factor. This keeps the thermal excess term and the thermal skewness calculated values both near zero for a Gaussian-like distribution and near unity for laminar flow.

**3b. First Derivative Thermal Moments**

Although the moments just derived work fine for defining the thermal boundary layer for turbulent boundary layers, we have found that there are some advantages to defining and using the first derivative-based moments as well. The proposition then, is that we can also describe the thermal boundary layer in terms of the central moments, $\lambda_n$, given by

$$\lambda_n = -\int_0^\infty dy\, (y - \delta_T^*)^n \, \frac{d\theta(y)}{dy} \quad . \tag{10}$$

where the thermal displacement thickness $\delta_T^*$ (Eq. 6) is the mean location. For this set of moments, it is convenient to define a set of auxiliary integrals defined as

$$\beta_n = \int_0^\infty dy\, y^n \, \theta(y) \quad . \tag{11}$$

These thermal displacement like integrals lets us calculate the higher order $\lambda_n$ and $\phi_n$ moments without having to differentiate the thermal profile. Using integration by parts, the first derivative thermal boundary layer width, $\sigma_q = \sqrt{\lambda_2}$ can be conveniently calculated as $\sigma_q = \sqrt{-(\delta_T^*)^2 + 2\beta_1}$. For the boundary layer thickness $\delta_q$, we find the four-sigma thickness, defined as $\delta_q = \delta_T^* + 4\sigma_q$, behaves similar to the 99% thickness for the temperature profile. The physical description of the first derivative based thermal profile using this alternative kernel can be expanded by defining a pair of shape parameters; the thermal boundary layer skewness, $\chi_q \equiv \lambda_3/\sigma_q^3$ and the excess, $\xi_q \equiv (\lambda_4/\sigma_q^4) - 3$. Using integration by parts, it is easily verified that $\lambda_3 = 2(\delta_T^*)^3 + 3\beta_2 - 6\delta_T^*\beta_1$ and that the fourth moment is given by $\lambda_4 = -3(\delta_T^*)^3 + 4\beta_3 - 12\delta_T^*\beta_2 + 12(\delta_T^*)^2\beta_1$.

## 4. Application to experimental data
### 4a. Laminar Flow

To demonstrate the utility of the new descriptive tools, experimental data for the laminar and turbulent boundary layer are examined. For the laminar boundary layer, it was not possible to find experimental wind tunnel data to demonstrate the new boundary layer moment method so instead we turned to the similarity solutions to the flow governing equations. For the laminar flow case, we start with the Pohlhausen[4] solution for laminar flow on a heated (cooled) flat plate. In Fig. 2 we presented a calculated profile using the method from Pohlhausen[4] for a Prandtl number value of $Pr$=0.7. In Fig. 3a we use the same solution approach to calculate thermal profiles for a range of Prandtl numbers. This figure, Fig. 3a, is a recreation of Fig. 12.9 from Schlichting.[6] In Fig. 3b we present the calculated moment based parameters describing the thermal boundary layer thickness and shape. Fig. 3c shows that the four-sigma thickness tracks the 99% thickness very well. In Figs. 3b and 3c the thicknesses $\zeta_T$, $\sigma_T$, $\delta_T$, and $\delta_{99}$ are all in units of $\sqrt{\nu_\infty x/u_\infty}$.

Note that the Pohlhausen[4] calculation in Figs. 2-3 is essentially a constant property method in that all of the physical properties are assumed constant. Even if the properties are evaluated at the film temperature (average temperature), Kakaç, Shah, and Aung[7] indicate that this approach is only accurate for temperature differences between the plate and free stream of <5 K for liquids and <50 K for gases. A more realistic approach to model laminar flow over a heated plate has been introduced by Weyburne.[8] It employs a variable property similarity approach. The original intent was to develop new heat transfer coefficient approximations for gases and liquids for forced laminar flow over a uniformly heated flat plate at zero incidence angle. The development is based on solving the variable property boundary layer equations using a variable a property similarity transform that incorporates an adjustable similarity scaling constant $\varepsilon$ such that $\eta^* = \varepsilon\eta$ (where $\eta$ is given by Eq. 2). The scaling constant's value $\varepsilon$ was iteratively adjusted until the scaled temperature gradient-at-the-wall value is equal to the Pohlhausen's[4] small temperature difference value and then this $\varepsilon$ value was approximated in terms of a simple function of the kinematic viscosity and the Prandtl number evaluated at the plate and free stream temperatures. The approximate scaling constant was then used to form new approximations for the heat transfer coefficients of gases and liquids. We include this short review because we want to emphasize even though the scaling constant $\varepsilon$ is fixed at unity ($\varepsilon = 1$), the calculations incorporate variable property behavior in order to realistically model laminar flow temperature profiles.

The variable property similarity approach is demonstrated in Fig. 4. In this figure the scaled temperature profiles for air flow are shown for temperature differences ranging from 1 K (solid line) to 350 K (dashed line). In Fig. 5 we plot the boundary layer thickness and shape for laminar air flow, water flow, and hydraulic fluid flow past a heated flat plate. As above, the thickness $\delta_T$ is in units of $\sqrt{\nu_\infty x/u_\infty}$. We should note that the variable property similarity solution requires an explicit fluid velocity $u_\infty$ value in order to calculate the Eckert number. For purposes herein, we choose velocities typical of what could be achieved in an experimental flow tunnel: 10 m/s for air, 5 m/s water, 0.5 m/s hydraulic fluid. Also note that for the calculations in Fig. 5, the fluid temperature $T_\infty$ was 300 K for air, 273 K for water, and 258 K for hydraulic fluid.

**4b. Turbulent Flow**

For the turbulent boundary layer, the number of available turbulent boundary layer datasets are limited. One of the most extensive sets available is from Blackwell, Kays, and Moffat.[9] In that study, an experimental investigation of the heat transfer behavior of the near equilibrium transpired turbulent boundary layer with adverse pressure gradient was carried out. Adverse pressure gradients of the form $u \propto x^m$, m = 0, -0.15, and -0.2, were studied along with a variety of transpired conditions.

In Fig. 6 we examine the effect of the adverse pressure gradient on the thermal boundary layer thickness, skewness, and excess for the non-transpired boundary layer (blowing, suction turned off) at different Reynolds numbers. The results for the thermal boundary layer thickness are somewhat unexpected; as the pressure gradient increases the thermal boundary layer thickness stays the same when plotted versus the Reynolds number based on the velocity momentum thickness. Notice that the skewness and excess results indicate that the shape of the profiles changes only slightly for the different pressure gradient cases. In order to make a visual correlation between the small shape changes we include Fig. 6d which shows three temperature profiles with m = 0, -0.15, and -0.2. The profiles correspond to the no blowing 091871, 120471, and 110871 datasets taken at x=70 in.[9]

A second set of temperature profiles are obtained from du Puits, Resagk and Thess.[10] In their study, turbulent thermal profiles were obtained in a Rayleigh–Bénard convection experiment that is a very close analog to the forced flow over a heated plate case at certain locations. The low Reynolds number thermal convection driven flow is generated in an enclosed chamber with a cooled top plate and a heated bottom plate for different height to width aspect ratios, $\Gamma$. When we examined the reported experimental data,[10] we found that the reported asymptotic values for the $\theta(z)$ profiles taken on the bottom plate differ from one by as much as 5%. (The z-direction in their experiments corresponds to the y-direction in Fig. 1). Apparently, some of the temperature drop between the hot and cold plates occurs in the bulk fluid. The moment method assumes the boundary layer asymptotic value is exactly one. Therefore, we created a new scaled profile, $\theta*(z)$, by dividing the $\theta(z)$ profiles values by the boundary layer edge value so that the profiles $\theta*(z)$ all asymptote to one. The scaled profiles for different height-width aspect ratios $\Gamma$, are shown in Fig. 7a versus $z/\delta_T^*$. The thermal displacement values, $\delta_T^*$, are calculated using the $\theta^*(z)$ values. It is apparent that the thermal profiles at different aspect ratios $\Gamma$ behave similar-like when scaled with $\delta_T^*$.[10] Note that the similar-like behavior of the profiles using $\delta_T^*$ as the scaling parameter has been predicted theoretically by Weyburne.[11]

Although the moment integrals are tolerant of some noise, we did find that noise at the boundary layer edge was a problem for the higher order moment calculations. This proved to be a problem for the Fig. 7a profiles. As a simple fix, it was decided to calculate the shape parameters $\chi_T \equiv \kappa_3/\sigma_T^3$ and $\xi_T \equiv (\kappa_4/\sigma_T^4)-3$ for a composite profile formed by averaging the profiles for each Ra number dataset. The resulting profile for Ra=5e10 is shown in Fig. 7b along with the calculated shape parameters $\chi_T$ and $\xi_T$ shown in the inset. We note that the values of $\chi_T \sim 1.6$ and $\xi_T \sim 2.2$ are very close to the low Reynolds number values for turbulent forced flow over a heated plate obtained by Blackwell, Kays, and Moffat[9] shown above (Figs. 6b and 6c).

The du Puits, Resagk and Thess[10] paper also included wall heat flux $q_w$ data that was measured using local embedded thermal sensors. Using Fourier's Law of heat conduction, this allows us to calculate the thermal diffusive parameters $\gamma_1$ and $\sigma_d$. In Table 1, we show some of this thermal diffusive data along with the four-sigma thickness $\delta_T$ using the $\kappa_n$ moment kernel. The moment method therefore permits us to track the turbulent boundary layers thermal profile thickness and shape and at the same time track the thickness and shape of the boundary layer region where thermal diffusivity plays a major role in the heat transfer process. Note that it is not necessary to actually differentiate the thermal profile to obtain the thermal diffusive parameters $\gamma_1$ and $\sigma_d$.

## 5. Numerical calculation of parameters

For the numerical calculation of the integrals reported herein, the Trapezoidal Rule was used and the data point $\theta(0)=1$ was added to every data set (when necessary). While the calculations are straightforward, one area which must be monitored is the integral value outside the boundary layer region (i.e. nominally in the free stream). In general, it was found that the reduced temperature $\theta$ for experimental data can fluctuate a few thousands about zero just above the boundary layer edge in the vicinity of the free stream. This small fluctuation resulted in nontrivial contributions to the calculated higher-order moments. To avoid this experimental noise problem, we rounded all the normalized temperatures $\theta$ that were nominally in the free stream to zero. The free stream starting point was taken as the first point at which the reduced temperature was less than or equal to zero. It is advisable to examine all moment integrands to insure the nominally free stream contributions are small compared to the boundary layer contributions.

## 6. Discussion

The new boundary layer moment method is based on the standard mathematical method for describing probability density functions. It is expected to work equally well on laminar and turbulent thermal boundary layers. The most important result of this approach is that for the first time, we have a mathematically well-defined way to describe the thermal boundary layer thickness and shape. Like the velocity boundary layer method described earlier,[2,3] the new boundary layer parameters each have a direct physical interpretation as to the thickness and shape of the profile (*e.g.* the mean location, the skewness).

Prior to this effort, the thermal boundary layer thickness standard measure was the $\delta_{99}$ parameter. Its use has persisted because there have been no good alternatives until now. For experimental datasets, the traditional $\delta_{99}$ thickness is usually calculated as the linear interpolation between two data points. This is because the actual functional form of the transition to the free stream for laminar and turbulent flow is unknown, hence it is not possible to fit for the $\delta_{99}$ thickness. Since experimental noise associated with each data point is sometimes on the order of a 1%, then the $\delta_{99}$ parameter for certain laminar and turbulent experimental datasets is simply ill defined. All one can do is to report a maximum and minimum value that brackets the probable value of $\delta_{99}$. In contrast, the new moment-based method provides a set of mathematically well-defined parameters based on integrals involving the whole dataset, not just a few noisy data points in the tail region.

Not only are the new parameters mathematically well defined, but we now have a set of parameters that truly help describe the shape of the profile in the thermal skewness and excess. Note that prior to the development herein, there have been no traditional shape parameters in regular use for the thermal profile, nothing like $H_{12}$ for the velocity profile. The skewness and excess provide a portrayal of the shape of the boundary layer profile. Whereas the skewness and excess for purely Gaussian curves are identically zero, this is not the case for the laminar and turbulent thermal profiles observed to date. For small temperature differences, the laminar flow thermal skewnesses and excesses are close to unity (see Fig. 3b). In contrast, for the turbulent boundary layer, the skewness values are also close to unity (Fig. 6b) whereas the excess values are closer to zero (Fig. 6c). Whether these observations are universally applicable will have to wait until more datasets are examined in the future.

Similar to the velocity profile development,[2,3] one of the interesting findings uncovered in this study is that the mean location of the first derivative of the thermal profile is the thermal displacement thickness $\delta_T^*$ given by Eq. 6. This means that $\delta_T^*$ is more than just the thermal displacement thickness in its usual sense but is also a true thickness parameter in its own right. It is a mean thermal location. In fact, Weyburne[11] has proven that if thermal profile similarity exists in a set of profiles then $\delta_T^*$ must be a similarity scaling parameter. Although we did not use the first derivative moments in the experimental section, we have confirmed that thermal boundary $\kappa_n$ and the $\lambda_n$ moments track each other for laminar and turbulent flows. One of the first derivative moment advantages is that the four-sigma thermal boundary layer thickness can be calculated with only $\delta_T^*$ and $\beta_1$ (Eq. 11) values.

The second derivatives moments are useful in determining the area of the boundary layer where thermal diffusive heat transfer is important. With the addition of the new moment kernels, we can now track the overall boundary layer thickness and the thermal diffusive thickness (see Table 1, for example). Note that it is not necessary to actually differentiate the thermal profile to obtain the thermal diffusive parameters $\gamma_1$ and $\sigma_d$. Fourier's Law of heat conduction permits the calculation of these parameters knowing the wall heat flux $q_w$ and the thermal displacement thickness. The wall heat flux $q_w$ can be measured using local embedded thermal sensors.[10] This is similar to the velocity profile case in which the second derivative based viscous boundary layer thickness can be calculated without differentiation by using the skin friction coefficient and the velocity displacement thickness.[3] One cautionary word on the thermal second derivative moments must be mentioned. For large temperature differences, the second derivative of the temperature, which normally has negative values, begins to take on significant positive values in the vicinity of the wall. This behavior becomes more pronounced as the temperature difference increases. In fact, the variable property similarity solutions for air flow in which the plate temperature is 200 K hotter than the free stream temperature results in a calculated value for $\sigma_d^2$ that actually becomes negative.[2] It is not clear whether this can be simply fixed by using the absolute value of the second derivative or if something else is necessary. More research is needed to explore this behavior.

## 7. Conclusion

In conclusion, a mathematically unique definition for the thermal boundary layer thickness has been developed along with a number of additional parameters useful in characterizing the shape of the thermal profiles of the boundary layer. For the turbulent boundary layer, the moment method allows one to track the thermal profiles thickness and shape and at the same time track the thickness and shape of the near wall region where thermal diffusivity plays a major role in the heat transfer process.

## Acknowledgement

This work was supported by the Air Force Office of Scientific Research, Gernot Pomrenke, Program Manager, and the Air Force Research Laboratory. In addition, the author thanks the various experimentalists for making their datasets available for inclusion in this manuscript.

**Figures and Tables**

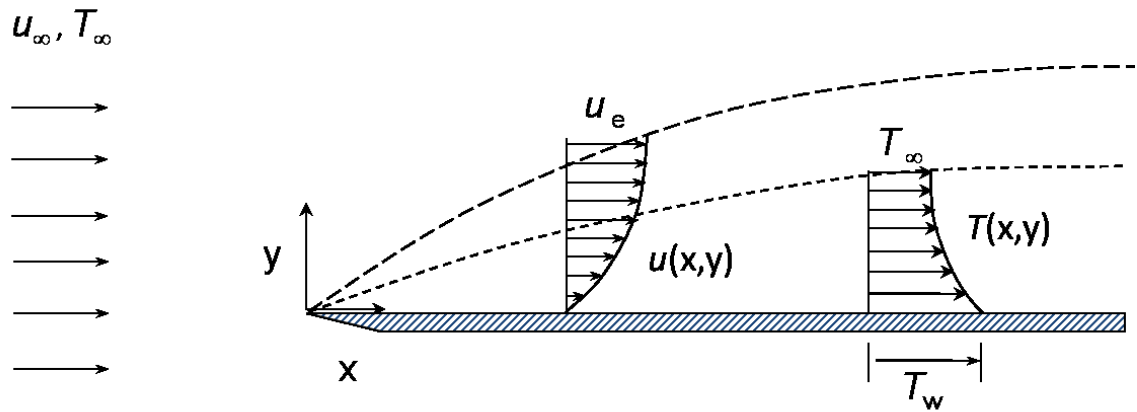

Fig. 1: A schematic diagram showing the flat plate 2-D flow geometry and variables.

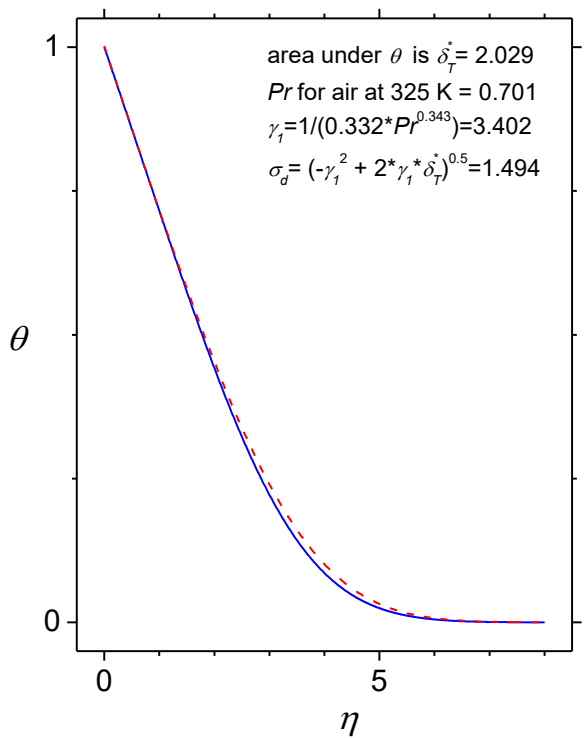

Fig. 2: Solid blue line is Pohlhausen's[4] calculated $\theta$ and the dashed red line is the Gaussian approximation (Eq. 1).

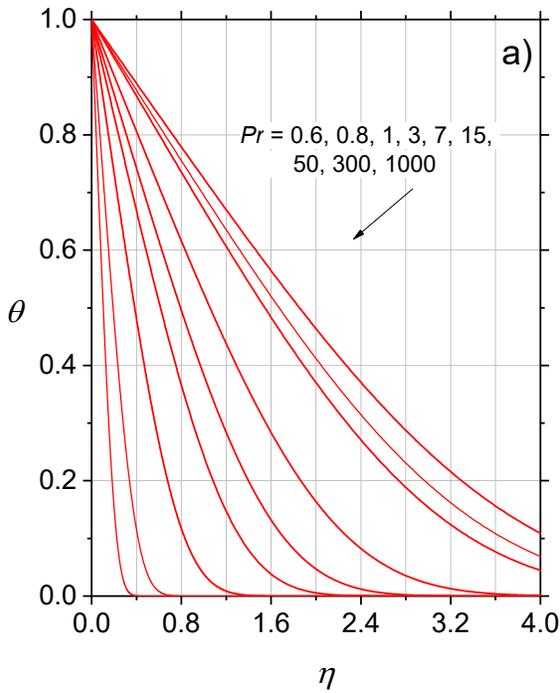

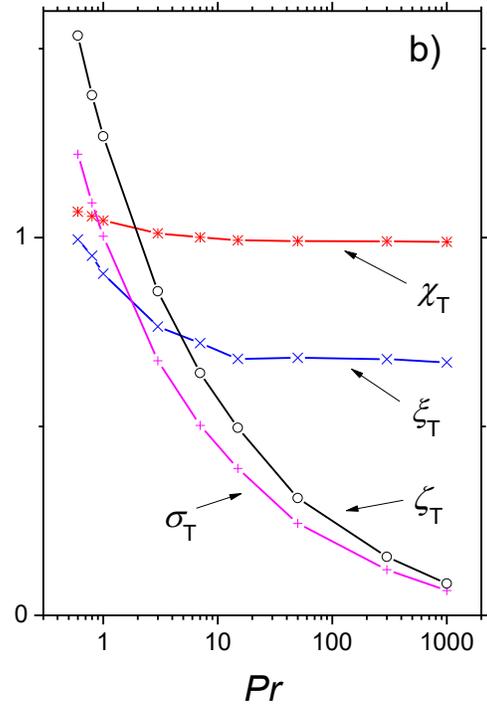

Fig. 3a: Laminar flow thermal profiles for a range of Pr numbers (after Schlichting[6]).

Fig. 3b: The mean location ($\zeta_T$), width $\sigma_T$ (+), skewness (∗), and excess (×) for the profiles in Fig. 3a (thickness units are $\sqrt{\nu_\infty x / u_\infty}$).

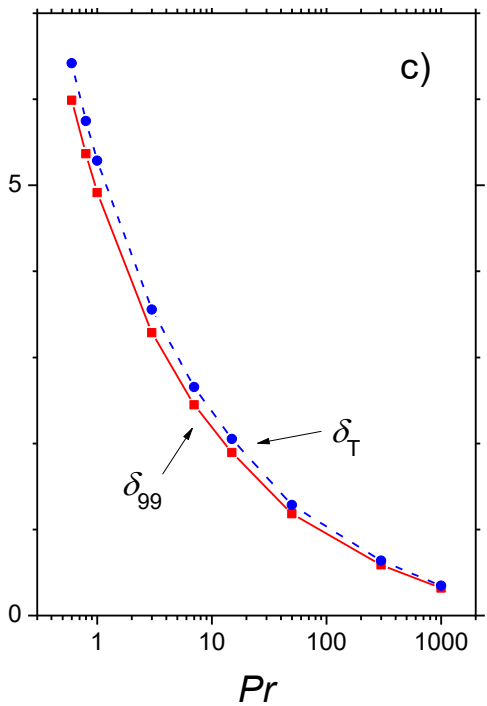

Fig. 3c: The thermal thickness $\delta_T$ (●) and $\delta_{99}$ (■) in units of $\sqrt{\nu_\infty x / u_\infty}$ for the profiles in Fig. 3a.

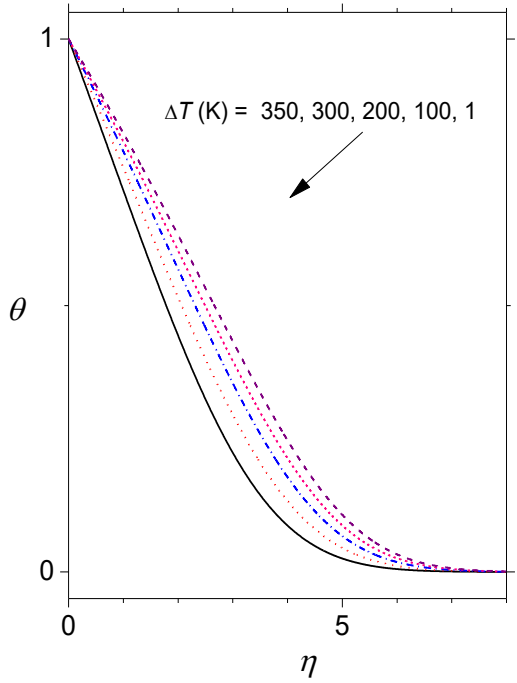

Fig. 4: The temperature profile for laminar air flow for temperature differences ranging from 1 K (solid line) to 350 K (dashed line).

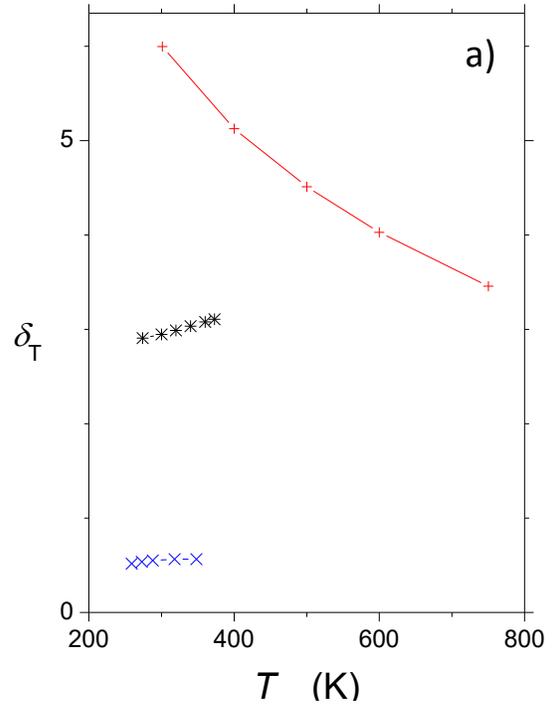

Fig. 5a: The laminar boundary layer thickness for air (+), water (∗), and hydraulic fluid (x) flow.

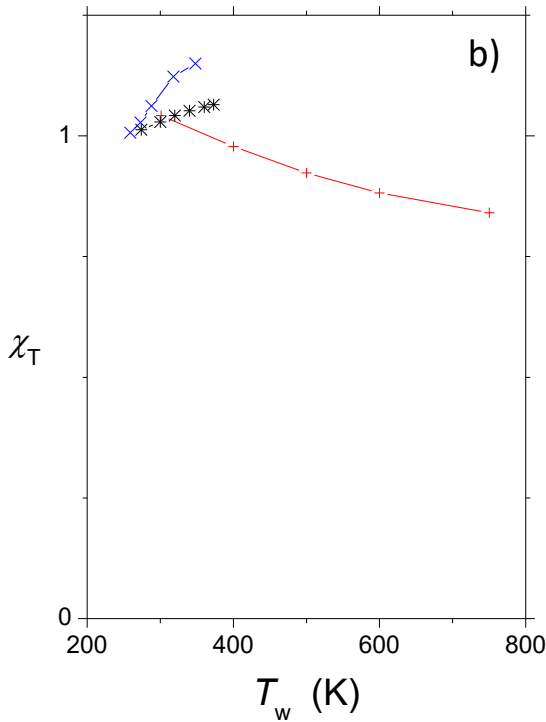

Fig. 5b: The laminar boundary layer thermal skewness for air (+), water (∗), and hydraulic fluid (x) flow.

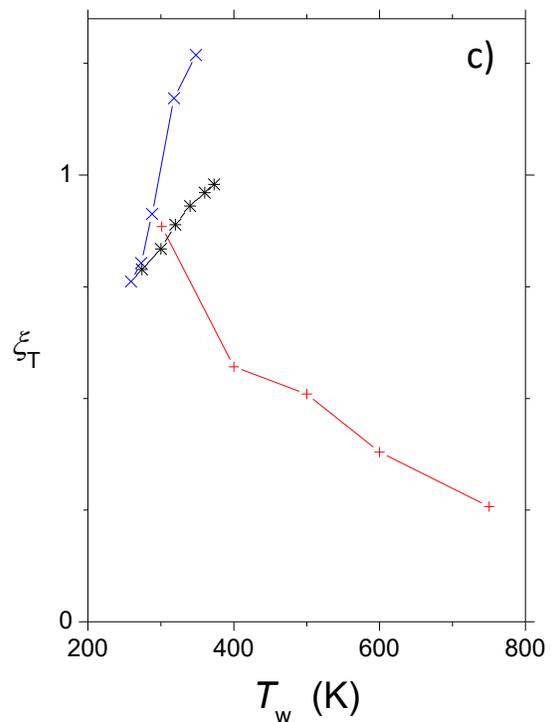

Fig. 5c: The laminar boundary layer thermal excess for air (+), water (∗), and hydraulic fluid (x) flow.

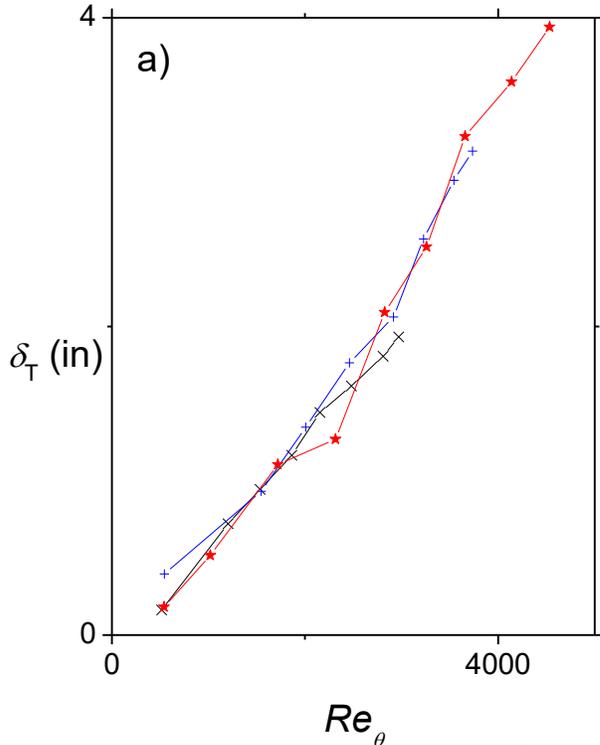
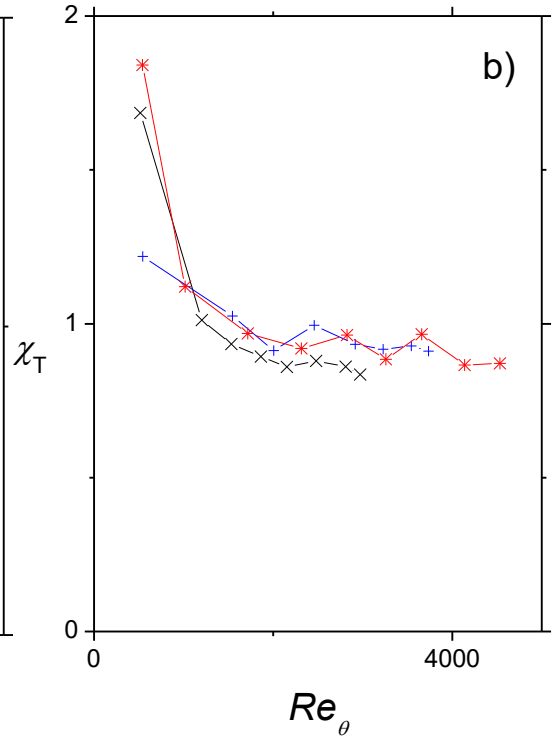

Fig. 6a: The turbulent boundary layer four sigma thermal thickness for m=0 (x), m= -0.15 (+), and m=-0.2 (*), data from Ref. 9.

Fig. 6b: The turbulent boundary layer thermal skewness for m=0 (x), m=-0.15 (+), and m=-0.2 (*), data from Ref. 9.

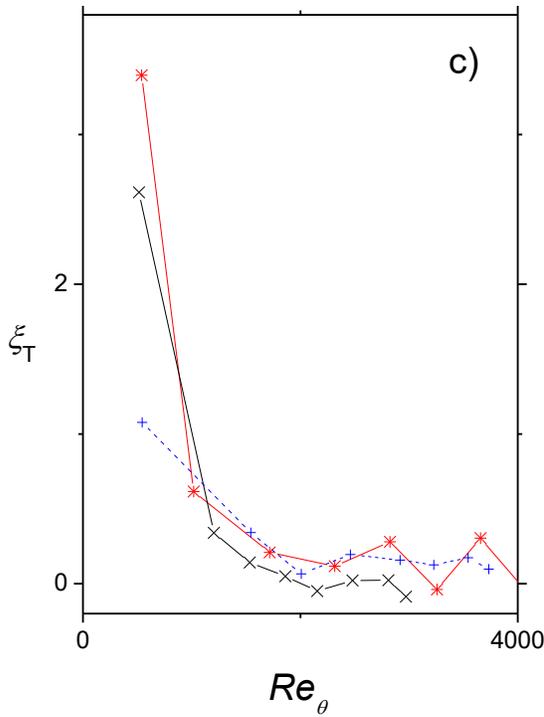
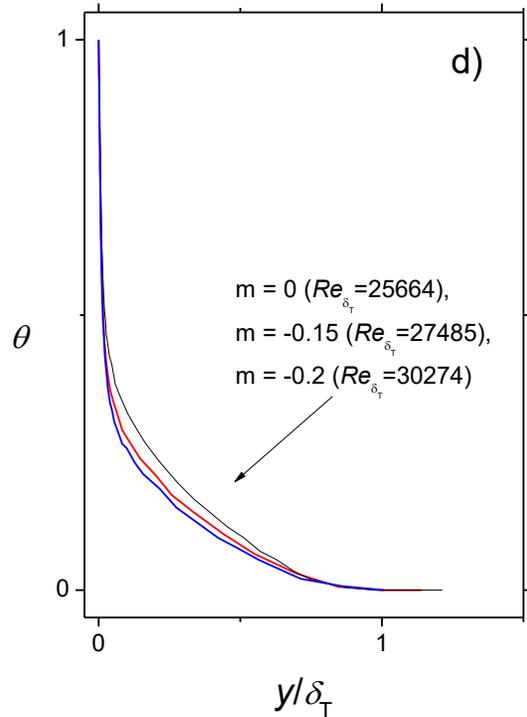

Fig. 6c: The turbulent boundary layer thermal excess for m=0 (x), m=-0.15 (+), and m=-0.2 (*), data from Ref. 9.

Fig. 6d: The turbulent boundary layer temperature profiles for m=0, m=-0.15, and m=-0.2, data from Ref. 9.

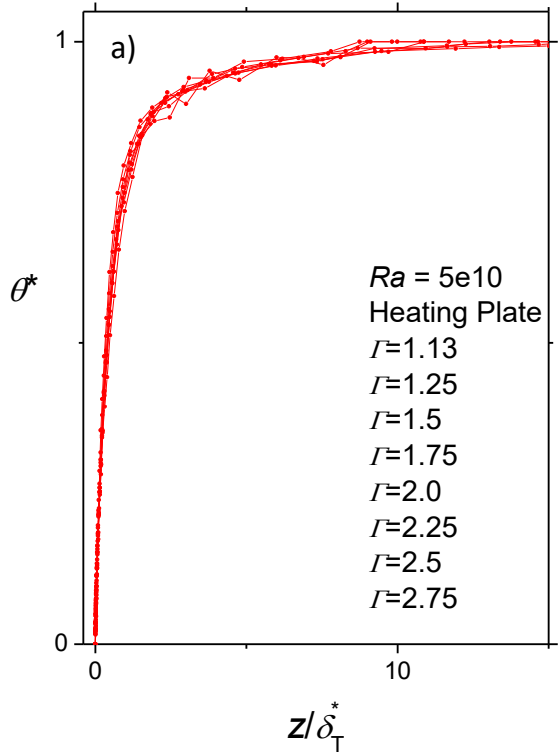 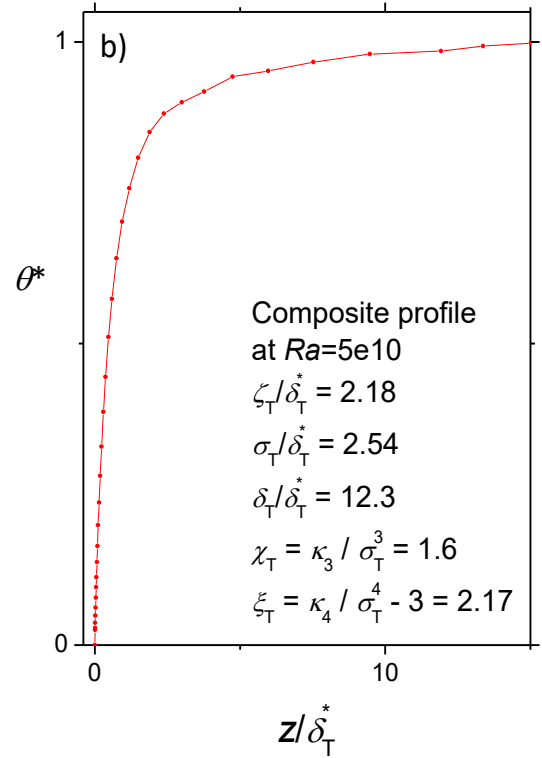

Fig. 7a: Eight turbulent boundary layer thermal profiles at different aspect ratios, data from du Puits, Resagk and Thess.[10]

Fig. 7b: Average of the eight profiles from Fig. 7a. This allowed for a good estimate of $\chi_T$ and $\xi_T$.

**Table**

| $\Gamma$ | Ra | q (W/m²) | Diffusivity Mean Location $\gamma_1$ (m) | Diffusivity Width $\sigma_d$ (m) | Diffusivity Thickness $\delta_d = \gamma_1 + 4\sigma_d$ | Boundary Layer Thickness $\delta_T$ (m) | Ratio $\delta_T/\delta_d$ |
|---|---|---|---|---|---|---|---|
| 1.13 | 5.2E10 | 9.3 | 0.0068 | 0.0134 | 0.0606 | 0.172 | 2.8 |
| 1.25 | 5.18E10 | 12 | 0.00701 | 0.0107 | 0.0498 | 0.104 | 2.1 |
| 1.5 | 5.25E10 | 22.5 | 0.00675 | 0.0098 | 0.0459 | 0.110 | 2.4 |
| 1.75 | 5.26E10 | 36.7 | 0.0064 | 0.0067 | 0.0332 | 0.076 | 2.3 |
| 2 | 5.26E10 | 55.7 | 0.00623 | 0.0087 | 0.0409 | 0.113 | 2.8 |
| 2.25 | 5.2E10 | 79.7 | 0.00613 | 0.0040 | 0.0220 | 0.054 | 2.5 |
| 2.5 | 5.19E10 | 114.5 | 0.00582 | 0.0034 | 0.01934 | 0.059 | 3.0 |
| 2.75 | 5.2E10 | 162.8 | 0.00542 | 0.0029 | 0.0169 | 0.058 | 3.5 |

Table 1: Thermal Boundary Layer Thickness Ratio as a function of aspect ratio $\Gamma$ for turbulent boundary layer datasets from du Puits, Resagk and Thess.[10]